\documentclass{ws-procs9x6}

\begin{document}

\title{Baryons from the lattice: past, present and future}

\author{Gunnar~S. Bali}

\address{Department of Physics \& Astronomy \\
The University of Glasgow,\\
University Avenue,\\
Glasgow G12 8QQ, UK\\
E-mail: g.bali@physics.gla.ac.uk}

\maketitle

\abstracts{I review recent lattice results on the spectrum and structure
of baryons. Limitations due to the quenched approximation and un-physically
heavy up and down quarks are discussed, and interfaces between
first principles studies of QCD (or approximations to it)
and ``model building'' are highlighted.}

\section{Introduction}
With new exciting experiments ongoing and planned at several
facilities, distributed over three continents, the past few years have
witnessed the emergence of a new arena of research on the borderline
between particle and nuclear physics: hadron physics.
These developments have been complemented by several break-throughs
in our theoretical understanding of the long distance, non-perturbative
regime of hadronic matter, directly from QCD. Here I will focus on
the most recent lattice results.

Progress in the field has been slow but steady.
Several ``revolutions''
in the lattice technology,
for instance non-perturbative
renormalisation and improvement programmes of action and operators,
were often hardly identifiable for the outside observer, a feature
that we share with many other fields of research.
In the case of strong QCD the failure of immediate satisfaction
is due to the
highly non-trivial phenomenology emerging from a Lagrangian
as simple as that of QCD and yet so colourful and complex that
even now it keeps
experimentalists and theoreticians very busy. 

A major but all to often neglected
application of lattice results is to establish validity range
and applicability of
phenomenological models or effective field theories directly from QCD
and to calculate low energy
parameters that appear in these models or approximations.
In quite a few situations, lattice predictions have matured to a degree
that allows direct confrontation with experiment.
Triggered by new experimental
programmes and improved interaction with other theoreticians,
an increasing number of lattice practitioners became interested
in what has now become
the domain of hadronic physics.
This more focused approach resulted in an accelerated development of
the field over the past two years with more to come in the future.
I will go through the present state of lattice studies, discuss the
two major sources of error, and review results on
spectrum and structure of baryons, in particular $N\gamma\rightarrow\Delta$
transition form factors, charge distributions, three-body forces and
the hyperon spin content.

\section{The state of the lattice}
Simulation results are obtained at finite lattice spacings and
have to be extrapolated to the continuum
limit. Moreover, the space-time box size should be big enough to
comfortably accommodate the hadronic scales of interest.
The dominant systematic uncertainties of present-day lattice results
however tend to be related to unrealistically heavy $u$ and
$d$ quark masses,
which typically
have to be {\em chirally} extrapolated over a long distance from
$m_{\pi}^2>0.25\,m_{\rho}^2$ to the {\em physical} limit
$m_{\pi}^2\approx 0.03\,m_{\rho}^2$, often combined with the
{\it quenched approximation},
i.e.\  neglecting the polarisation effects
of sea quarks on the QCD vacuum.

The latter approximation, which easily
reduces the computational effort by a factor of $10^3$, is justified in
the limit of the number of colours $N_c\rightarrow\infty$, however,
it is not {\em a priori} clear whether $1/3\ll 1$. Phenomenology\cite{Lebed}
suggests that this assumption is not as outrageous as it might sound
at first. This seems substantiated by simulations of pure
$SU(N_c)$ gauge theories
for\cite{Lucini:2002ku} $N_c=2, \ldots,6$  as well as by a comparison
of quenched lattice predictions of the light hadron spectrum with
experiment:
ratios of light hadron masses on the lattice
indeed have at last been found to be inconsistent with
the observed spectrum.\cite{Aoki:2002fd}
The differences are typically smaller than
10~\%.

However, there are cases in which ``quenching'' clearly matters.
Quantum mechanical perturbation theory tells us
that an ${\mathcal O}(\epsilon^2)$ correction to an energy eigenvalue
corresponds to an ${\mathcal O}(\epsilon)$ correction
on wave functions which hence
can easily be changed by as much as 30~\%. This will affect quantities like
decay constants and parton distributions.
The situation with for instance bottomonium fine structure
splittings or electroweak decay rates is even worse,\cite{Bali:1998pi}
as these are roughly
proportional to the square of the wave function at the origin.
Strong decays do not exist in this approximation as quarks and anti-quarks
cannot annihilate, and therefore one might hesitate to expect very broad
resonances to be
modelled in a reasonable way. The same also holds true for any
valence quark based model. We shall discuss some other shortcomings of
{\em quenching} below.

Fortunately, with several new
supercomputer installations on the horizon
by early 2004 we are at the brink of the {\em fourth generation} of
simulations including sea quarks. These certainly will bring us down
in the quark mass to about the point where the $\rho$ meson becomes
unstable and will at the same time allow for controlled continuum limit
extrapolations.

\begin{figure}[t]
\centerline{\epsfxsize=3.5in\epsfbox{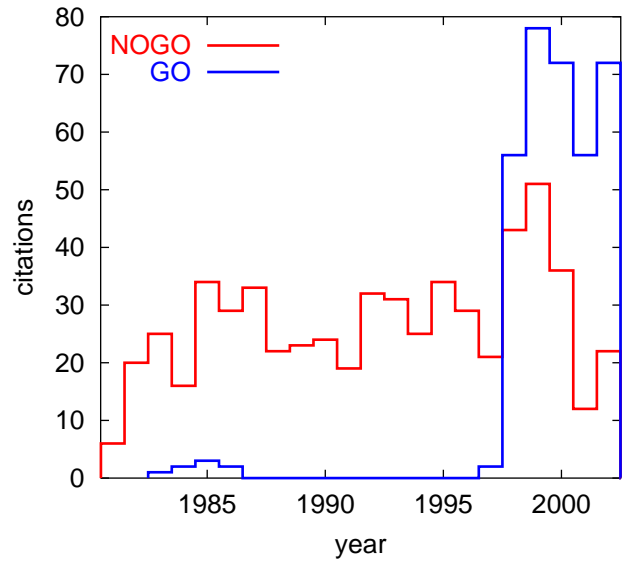}}
\caption{Yearly SPIRES citations of Nielsen-Ninomya\protect\cite{Nielsen:1980rz} (NOGO)
and GW\protect\cite{Ginsparg:1981bj} (GO).\label{gwcite}}
\end{figure}

Recently a way has been found to implement the analogue of
exact {\em chiral symmetry} without ``doublers'' on the lattice,
known as overlap,
Neuberger or Ginsparg-Wilson (GW) fermions.\cite{Neuberger:1997fp}
One particular
(truncated) realisation of the Neuberger operator are domain wall
fermions.\cite{Kaplan:1992bt}
GW fermions are the
only known consistent way of regularising a quantum field theory with
massless fermions (like e.g.\ the standard model) and hence obviously
a major theoretical
break-through in particle physics as a whole. The famous
Nielsen-Ninomiya no-go theorem\cite{Nielsen:1980rz} is circumvented by
relaxing the definition of {\em chiral symmetry}.\cite{Ginsparg:1981bj}
I illustrate the yearly number
of citations\cite{Hasenfratz}
of these two important articles\cite{Nielsen:1980rz,Ginsparg:1981bj}
published in 1981 and 1982, respectively, in Figure~\ref{gwcite}
and leave judgement to the reader
which paper would have better impacted on Research Assessment
Exercises.
In practical terms GW fermions mean
that realistically light quark masses will become accessible
to future numerical simulations, first quenched\cite{Lee:2002gn} and
subsequently un-quenched.

Going to light quark masses is expensive as the linear spatial
lattice extent $L$ has to be large relative to the inverse pion mass,
to avoid strong finite size effects from pions exchanged ``around''
the boundaries. For instance
if we intend to simulate
a physical pion with say $L>4/m_{\pi}$ this implies $L>5.7$~fm.
The main limitation however is
that the effort of inverting the lattice Dirac operator
scales with a large power of the inverse quark mass.
This will improve with GW quarks that do not suffer from
un-physical zero or near-zero modes: while at $m_{\pi}>0.5\,m_{\rho}$
GW fermions are easily two orders of magnitude more expensive to simulate
than ``standard'' formulations, there will be a reward at small quark masses.
In the quenched approximation this improvement factor has already turned
out to be infinite, rendering a pion lighter than 200~MeV
from impossible to possible.\cite{Lee:2002gn} Chiral fermions
also greatly simplify
operator mixing and renormalisation.

\section{Chiral extrapolations and ``quenching''}
Based on the assumptions that QCD bound states are mesons and
baryons, that there is a mass gap and spontaneous chiral
symmetry breaking at zero quark mass, an effective low energy chiral
Lagrangian can be derived in the spirit of the
Born-Oppenheimer approximation.
This Lagrangian will, to leading order,
describe interactions between the (fast moving)
massless Nambu-Goldstone pions and
other hadrons. In nature quarks and thus pions are not massless
and the leading mass corrections are formally
of order $m_{\pi}/\Lambda_{\chi SB}$ where $\Lambda_{\chi SB}\approx
4\pi F_{\pi}>1$~GeV. Of course in an effective field theory the
number of terms explodes at higher orders and predictive power is eventually
lost,
unless an early truncation is possible.\cite{Nieves}

{\em Third generation} lattice simulations with sea quarks were limited
to unrealistically heavy
pions, heavier than about 400~MeV. Only
recently masses as low as 180~MeV
have become possible in the quenched approximation,\cite{Lee:2002gn}
albeit not yet at sufficiently large volumes. To allow for a
first principles connection with phenomenology it is mandatory
to establish an overlap region between lattice simulations
and (leading order) chiral perturbation theory.
In general the size of this window will depend on the observable in question.
GW fermions help a lot since only in this case we
have a variant of
chiral symmetry at finite lattice spacings while strictly speaking
in other formulations (without doublers)
a sensitive comparison can only be performed
after extrapolation of lattice results to the continuum limit.
While even $400$~MeV~$\ll\Lambda_{\chi SB}$ (modulo the ambiguity of ``$\ll$''
 {\em vs.}\/
``$<$'') such a pion is still doomed to ``see'' some of the internal
structure of the proton, with an inverse charge radius of about 250~MeV:
it is doubtful that the quark and gluon
nature of QCD can completely be ignored with such a ``hard'' pion probe.
With broadening baryonic wave functions such issues are likely to
be more critical in QCD with light sea quarks than they already are
in the quenched approximation.

\begin{figure}[t]
\centerline{\rotatebox{90}{\epsfysize=3.5in\epsfbox{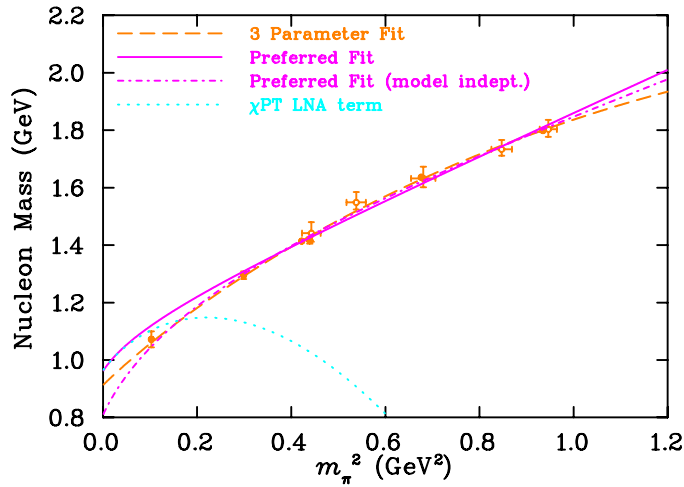}}}
\caption{Chiral extrapolation of the nucleon mass, after phenomenological
corrections for finite lattice spacing and volume
effects.\protect\cite{Young:2002cj} The data are from the
CP-PACS\protect\cite{AliKhan:2001tx} (solid circles)
and UKQCD Collaborations\protect\cite{Allton:2001sk} (open circles).\label{chiext}}
\end{figure}

Na\"{\i}vely, hadron masses are a polynomial in
the quark mass, $m_q\propto m_{\pi}^2$. However, pion
loops give rise to non analytic (NA) functional dependence on the quark
mass. For instance the nucleon
mass is given by,
\begin{equation}
\label{eq:fit}
m_N(m_{\pi})=m_N(0)+c_2m_{\pi}^2+c_3m_{\pi}^3+\cdots.
\end{equation}
The coefficients of the NA terms are related
to phenomenological 
low energy constants. For instance $c_3=-3g_A^2/(32\pi F_{\pi}^2)$.
In the quenched approximation, which neglects disconnected fermion lines,
the leading NA (LNA) term is not proportional
to $m_{\pi}^3$ anymore but to $m_{\pi}^2\ln m_{\pi}$, known as a
{\em chiral log}. Only very recently
convincing signs of such logarithmic behaviour have been found, albeit
not yet with the expected coefficients.

Recently, corrections to standard LNA terms due to massive
pions (resulting in a different pion cloud) have been modelled,
in particular by the Adelaide group.\cite{Young:2002cj}
Unlike the chiral Lagrangian itself, such models cannot be derived in any
limit directly from QCD, and the predictive power has to
be digested with some caution. Figure~\ref{chiext} reveals that we have
not yet made contact with chiral perturbation theory (dotted curve,
$c_3$ fixed).
Nonetheless the data are well fitted by Eq.~(\ref{eq:fit}),
with $c_3$ as a free fit parameter (dashed curve),
and the extrapolated value has dangerously tiny
statistical errors. The Adelaide cloudy bag model (CBM) suggests a different
functional form that screens the LNA term at pion masses close to
a bag parameter, $\Lambda$.
With a fixed phenomenological value of $\Lambda$
the solid curve is obtained while a three parameter fit yields the
dashed-dotted curve. By treating the difference as a systematic
uncertainty such models allow us to arrive at more realistic error
estimates. Alternatives to the CBM include
NJL based\cite{Kvinikhidze:2001xb}
as well as Bethe-Salpeter based\cite{Bicudo:2001cg} ideas.

The quenched approximation is sick in several ways. I already mentioned that
many unstable particles become stable. It is not always entirely
clear when this ``bug'' is a virtue and in which cases not. A more serious flaw
is that unitarity is violated since quarks are not included 
in a 
consistent way. This defect, which becomes
visible in particular at small quark masses and in the scalar sector,
also reflects onto and can be understood in terms of the chiral expansion.
For instance chiral logs originate from this source.
Another related problem is the absence of the axial anomaly since
quarks do not feed back onto the glue. This leads to the spontaneous breaking
of a greater flavour symmetry group to start with
and results in $n_F^2$ instead of $n_F^2-1$
pions: the $\eta'$ becomes heavy only at small sea
quark masses,\cite{Bali:2001gk}
with possible implications onto the phenomenology of excited baryon states:
the mass splitting between the nucleon ($N=N^{1/2^+}$)
and the $N^*=N^{1/2^-}$ for instance is a consequence of
chiral symmetry breaking and its
size is related to transitions $N^*\rightarrow N+\pi\rightarrow
N^*$. The number of pions should  naturally be expected to
seriously affect such observables. (Note that also in the
quenched approximation
there is a pion cloud since valence
quarks can travel forwards and backwards and forwards again in time.
Pion exchange is possible too, as valence quarks are swapped.)

To this end it has recently been suggested to model
{\em quenched} LNA terms and to subtract these
from the corresponding lattice
simulation results, replacing them
by the expected {\em un-quenched} counterparts.\cite{Thomas:2002sj}
Needless to say that
many assumptions are required, in particular
relations between low energy parameters appearing in quenched and un-quenched
chiral Lagrangians and in the CBM. However,
in the absence of lattice results with light sea quarks, playing around
with such models provides us with valuable rough ideas of possible sizes
of quenching effects. On the other hand, once quality un-quenched data
become available this will positively feed back onto such models, providing
insights impossible to gain from experiment alone where quark masses and
the number of active flavours are fixed.

Determinations of nucleon parton distributions
seem to suffer {}from extremely large uncertainties in the chiral
extrapolation; in particular this is so for the leading moments. In a recent
study of the LHPC/SESAM  Collaborations\cite{Dolgov:2002zm}
of quark distributions including sea
quarks almost perfect agreement
was found with quenched results, obtained both by them and by
QCDSF,\cite{roger} down to
$m_{\pi}\approx 600$~MeV. However, a na\"{\i}ve polynomial extrapolation
of $\langle x\rangle_{u-d}$ was found to overestimate
the phenomenological value by
as much as a factor 5/3. Such a factor can indeed be
accounted for by modelling pion mass effects on
the LNA contribution
in a sensible way,\cite{Detmold:2001jb} indicating that the quark mass
is the dominant (and a huge) source of uncertainty.

\section{Spectrum}
In the past two years there has been a vibrant flow of lattice publications
on the spectrum of excited nucleon states.\cite{Richards}
All but one\cite{Maynard} have been based on the quenched approximation so far.
Only the LHPC/UKQCD/QCDSF Collaborations,\cite{Gockeler:2001db}
using the Wilson-Clover action,
attempted a continuum limit extrapolation.
Other strategies were implementations of improved actions
like the $D_{234}$ action by Lee {\it et al.}\cite{Lee:2001ts}
or the FLIC action by Melnitchouk {\it et al.}\cite{Melnitchouk:2002eg}.
The Riken-BNL group employed domain wall fermions to improve
the chiral properties.\cite{Sasaki:2001nf}

One feature that all these articles share is a level inversion
between the positive parity
Roper $N'(1440)$ resonance and the negative parity $N^*(1535)$,
relative to experiment. It was not clear whether this was
a quenching effect, hence shared with many models, most of which are
essentially {\em quenched} too. But maybe the resonance observed in nature
did something funny, to which the quark-only operator used on the
lattice to create the Roper did not couple well. A pessimist might also
argue that such resonances are quite unstable which could very well
affect their position in ways that are not easily foreseeable, prior to
doing the {\em real thing}.

\begin{figure}[th]
\centerline{{\epsfxsize=3.5in\epsfbox{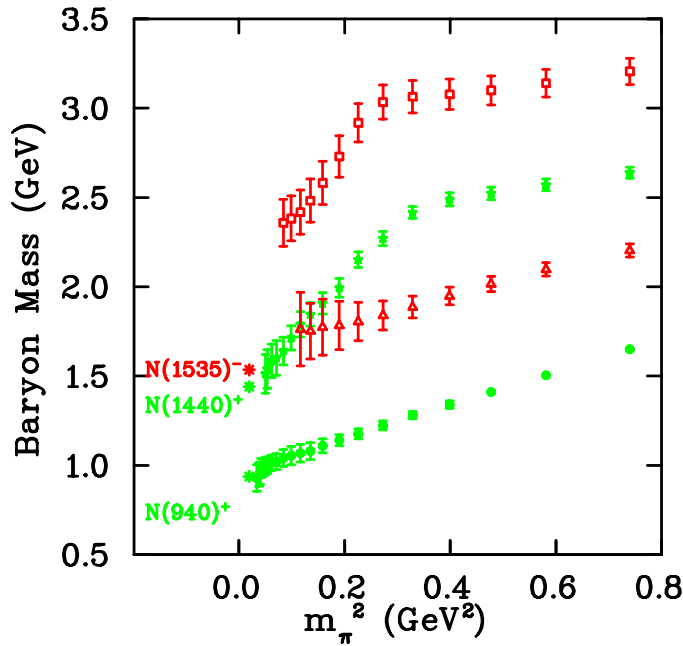}}}
\caption{Chiral extrapolation of the $N$, $N'$ and $N^*$
masses.\protect\cite{Lee:2002gn}\label{fig:lee}}
\end{figure}

However, new results obtained with
GW fermions by the Kentucky-Washingon group\cite{Lee:2002gn}
at lighter pion masses, $m_{\pi}>180$~MeV,
than ever realised before indicate an exciting behaviour: as the pion becomes
lighter than about 550~MeV the slope of the Roper data bends downwards,
heading straight in the direction of
the experimental value (cf.~Figure~\ref{fig:lee}).
While at the lightest
quark mass $Lm_{\pi}\approx 2.7$ and finite size effects are likely,
everything with $m_{\pi}^2>0.1\,\mbox{GeV}^2$ is pretty safe.
There is overlap with other lattice
studies\cite{Gockeler:2001db,Melnitchouk:2002eg,Sasaki:2001nf}
in the region
$m_{\pi}^2>0.25\,\mbox{GeV}^2$ and one might wonder why the onset of
this behaviour has not been detected before. Inspection shows, however, that
within the statistical errors an effect of the expected size would
not have appeared significant in these cases.

Certainly, once the independence of the Kentucky results from
Bayesian fitting priors and lattice volume has been firmly established,
these are very exciting news that tell model builders how transitions
between heavy and light quark regions can look like: a prime
example for the cross fertilisation between lattice QCD, models
and experiment in this thriving
area of research. The Kentucky group has investigated many
additional baryonic channels and it is
astonishing how good quenched QCD seems to
explain the positions of these strongly decaying resonances, once the quark
masses become sufficiently light, another unsolved miracle of nature.

\section{Structure}
Exciting results on the transition matrix element
for electromagnetic $N$ to $\Delta$ transitions have been reported.
The relevant operator
is usually decomposed into three Sachs type form factors, accompanying magnetic
dipole transitions, $G_{M1}(q^2)$ as well as electric and Coulomb/scalar
quadrupole transitions, $G_{E2}(q^2)$ and $G_{C2}(q^2)$.
The quadrupole transitions ratios
$R_{EM}=G_{E2}/G_{M1}$ and $R_{SM}=G_{C2}/G_{M1}$
are related to the question of nucleon deformation.\cite{Drechsel} 
Recent experimental results, in particular from the CLAS\cite{Joo:2001tw}
and OOPS Collaborations\cite{Mertz:1999hp} indicate the ratios
$R_{EM}\approx -0.02$
and $R_{SM}\approx -0.07$ around $q^2\approx 0.53$, which is the lowest
momentum transfer that could be realised in a recent lattice study with
two flavours of sea quarks, not much lighter than the
strange.\cite{Alexandrou:2002pw} These first quantitative lattice
results seem to get the sign right and point towards
values $R_{EM}=-0.03(1)$ and $R_{SM}=-0.03(2)$, consistent with experiment.
Interestingly, the quenched reference result $R_{EM}=-0.009(8)$ is
compatible with zero, indicating that a more realistic pion cloud might
be essential
for the effect. Obviously, lattice simulations are still miles
away from computing, say, octupole form factors for transitions between
orbitally excited nucleons.

In another lattice investigation\cite{Alexandrou:2002nn}
the electric charge and matter distributions
of $\pi$, $\rho$, $N$
and $\Delta$ have been studied at rest. One can easily calculate
the charge radius from such a distribution and
the Fourier transform is related to form factors at zero momentum
transfer.
While the
(cut-off and scheme dependent) matter distributions of all investigated
hadrons were found to be remarkably similar there are differences
in the charge distributions. In the quenched approximation
the pion is only approximately half as wide as the $\Delta$,
with $N$ and $\rho$ almost of the same size as the $\Delta$.
Even with
a pion mass of almost 600 MeV un-quenching significantly broadened the
latter three distributions, and in particular the $\Delta$,
while the $\pi$ remains completely unaffected. The quenched proton
charge radius comes out
to be $r_p=0.59(4)$~fm, after polynomial
extrapolation to the physical limit.
This has to be compared with the value
$r_p\approx 0.81$~fm from dipole fits to experimentally determined
electromagnetic form factors.
The above {\em direct} lattice result is in agreement with
$r_p\approx$~0.6~fm obtained {\em indirectly} from a dipole fit
to electric and magnetic
proton form factors calculated on the lattice
by the QCDSF Collaboration.\cite{Gockeler:2001us} However,
from recent JLAB Hall A results,\cite{Jones:uu}
showing an energy dependence of the ratio
$G_e/G_m$, we know that the dipole ansatz is not the whole truth,
which stresses the importance of determining quantities like
charge radii in more than one way.
The same QCDSF work suggests $r_p\approx 0.7$~fm
for two flavours of sea quarks, closer to the phenomenological value.

For the first time deformations of the hadron wave functions have been
investigated.\cite{Alexandrou:2002nn}
Obviously, a spectroscopic quadrupole moment becomes only possible
for an angular momentum of at least one,
such that deviations from
spherical symmetry are  ruled out for the nucleon. However, the
$\rho$ was reported to be somewhat prolate with the $\Delta$ being
slightly
oblate and the deformations appeared to increase when including sea quarks.
In neither of the cases was the deformation of the $\Delta$ statistically
significant though.

It is not clear whether QCD forces are {\em stringy}
with a three body potential that depends on the length of the
shortest path connecting the three quarks to one point (Mercedes or $Y$ law)
or if they merely can be understood in terms of sums of two body interactions
($\Delta$ law).\cite{Bali:2000gf} Leaving aside the question
whether an instantaneous Born-Oppenheimer interaction picture is
appropriate for a situation with light quarks, the nucleon wave function seems
to be better described
by a Schr\"odinger equation with a potential that depends on the
``$\Delta$-distance'' than by the $Y$ configuration.\cite{Alexandrou:2002nn}
This finding is also supported by recent lattice calculations of $SU(3)$ and
$SU(4)$ gauge theory as well as of the $Z_3$ Potts
model.\cite{Alexandrou:2002sn} The results
indicate that for interquark distances of physical relevance, i.e.\ $r<1$~fm,
the $\Delta$-picture yields a very accurate description while,
long after the ``string'' of QCD with sea quarks is broken,\cite{}
the $Y$ picture is
eventually approached.

Recently, spin and transversity of the
quark contributions to the $\Lambda$ hyperon have been calculated
by QCDSF.\cite{Gockeler:2002uh}
The fraction $\Delta q_{\Lambda}$
of the spin carried by quarks of flavour $q$
is given by the forward matrix element of the axial vector current,
$\langle \Lambda(p',s)|\bar{q}\gamma_{\mu}\gamma_{5}q|\Lambda(p,s)\rangle
=2\left(p'-p\right)_{\mu}\Delta q_{\Lambda},$
i.e.\ by the lowest moment of the helicity distribution.
Assuming $SU(3)$ flavour symmetry one can express the
$\Delta q_{\Lambda}$s in terms of proton spin fractions
$\Delta q_p$. Converting the lattice results into the $\overline{MS}$
scheme at $\mu=2$~GeV yields an
axial charge
$g_A=\Delta u_p-\Delta d_p$ that falls short of the experimental
value $g_A\approx 1.26$ by 10-20~\%, due to
quenching and/or due to uncertainties in the chiral extrapolation,
such that we cannot expect the result $\Delta u_{\Lambda}=\Delta d_{\Lambda}
=-0.02(4)$ to be consistent with the real world either.

The lattice data on proton
and $\Lambda$ spin fractions turn out to be in excellent agreement
with $SU(3)$ flavour symmetry, an observation that might be independent
of sea quarks and certainly
consistent with the success in
understanding semileptonic decay rates of the $\Lambda$.
Combining experimental data obtained for the proton helicity
with the systematic error on $SU(3)_F$ symmetry suggested by this lattice
study, yields the predictions $\Delta u_{\Lambda}=\Delta d_{\Lambda}=-0.17(3)$
and $\Delta s_{\Lambda}=0.63(3)$, again in the $\overline{MS}$ scheme
at $\mu=2$~GeV.

\section{Outlook}
Experimental efforts are paralleled by lattice studies of baryon
spectrum and structure. The main limitation that has to be overcome in
the near future is the use of unrealistically heavy quarks in quenched
simulations, as well as in simulations with sea quarks. Recent theoretical
developments and advances in computer technology mean that optimism
is justified.

\section*{Acknowledgments}
\uppercase{T}his work is supported by \uppercase{PPARC} grants \uppercase{PPA/A/S/}2000/00271, \uppercase{PPA/G/O/}1998/00559 and
\uppercase{EU} grant \uppercase{HPRN-CT}-2000-00145.

\end{document}